# The Japanese Smart Grid
# Initiatives, Investments, and Collaborations

Amy Poh Ai Ling
Graduate School of Advanced
Mathematical Sciences
Meiji University,
Kanagawa-Ken, Japan.

Sugihara Kokichi
Graduate School of Advanced
Mathematical Sciences
Meiji University,
Kanagawa-Ken, Japan.

Mukaidono Masao
Computer Science Department
School of Science and Technology,
Meiji University,
Kanagawa-Ken, Japan

*Abstract*—**A smart grid delivers power around the country and has an intelligent monitoring system, which not only keeps track of all the energy coming in from diverse sources but also can detect where energy is needed through a two-way communication system that collects data about how and when consumers use power. It is safer in many ways, compared with the current one-directional power supply system that seems susceptible to either sabotage or natural disasters, including being more resistant to attack and power outages. In such an autonomic and advanced-grid environment, investing in a pilot study and knowing the nation's readiness to adopt a smart grid absolves the government of complex intervention from any failure to bring Japan into the autonomic-grid environment. This paper looks closely into the concept of the Japanese government's 'go green' effort, the objective of which is to make Japan a leading nation in environmental and energy sustainability through green innovation, such as creating a low-carbon society and embracing the natural grid community. This paper paints a clearer conceptual picture of how Japan's smart grid effort compares with that of the US. The structure of Japan's energy sources is describe including its major power generation plants, photovoltaic power generation development, and a comparison of energy sources between Japan and the US. Japan's smart community initiatives are also highlighted, illustrating the Japanese government planned social security system, which focuses on a regional energy management system and lifestyle changes under such an energy supply structure. This paper also discusses Japan's involvement in smart grid pilot projects for development and investment, and its aim of obtaining successful outcomes. Engagement in the pilot projects is undertaken in conjunction with Japan's attempt to implement a fully smart grid city in the near future. In addition, major smart grid awareness activities promotion bodies in Japan are discuss in this paper because of their important initiatives for influencing and shaping policy, architecture, standards, and traditional utility operations. Implementing a smart grid will not happen quickly, because when Japan does adopt one, it will continue to undergo transformation and be updated to support new technologies and functionality.**

*Keywords*- *Japanese Smart Grid, Initiative, Investment, Collaboration, Low Carbon Emission*

## I. INTRODUCTION

Culture encroachment happens through the interplay of technology and everyday life. The emergence of the smart grid creates a drastic increase in the demand for the smart supply of energy flows [1]. The Japanese version of the 'smart city' is envisaged for the post-fossil fuel world. Alternative energy sources, such as solar, wind, and nuclear power, are harnessed in mass quantities [2]. In Japan, 'smart grid' implies energy transmission and distribution to promote the stability of the electric power supply, by using information and communication technology while introducing a high level of renewable energy [3]. The focus will be on how to stabilize power supplies nationwide as large amounts of wind and solar power start entering the grid. This is because, unlike conventional power sources, such as hydro, thermal, and nuclear power, solar and wind energies are prone to the vagaries of the weather. People in Japan are still not familiar with the smart grid concept because the system has yet to gain currency. According to a nationwide survey released in December 2010 by the advertising agency Hakuhodo Inc., only 36.4% of about 400 respondents aged from 20 to 70 years said they understood, or had heard of, a smart grid [4].

To address the likely impact of the smart grid on customers, utilities, and society as a whole, it may be necessary to conduct a pilot study [5]. It is widely understood that the new services enabled by the smart grid will include different rate designs that encourage curtailment of peak loads and make more efficient use of energy. The future grid will be an autonomic environment that helps users to not only share large-scale resources and accomplish collaborative tasks but also to self-manage, hence reducing user interventions as much as possible [6]. This paper discusses the Japanese concept of the smart grid and the differences between the smart grid movements in Japan and the US. The Japanese government aims to achieve a low-carbon society, employing the natural grid as a major source of energy supply for the country, whereas the US is focusing on its business and infrastructure development. This paper mainly focuses on Japan's smart grid structure, discusses Japan's initiatives, investments, and collaborations, and examines how the full implementation of the smart grid is anticipated to come into play comparing to an early written paper referring to the natural grid whereby only the smart community is illustrated [7].

## II. METHODOLOGY

Conceptual analysis is adopted as our methodology with the application of the hermeneutic circle. The hermeneutic circle is used for interpretive reasons [8] and because it enables a conceptual–analytical research method. Philosophers and theologians in reviewing something that is not explicitly present commonly apply this approach. This study investigates the assumptions of different Japanese smart grid initiatives, investments, and collaborations, for which the hermeneutic circle is a natural choice of research methodology. The methodology refers to the idea that what one understands of the text as a whole is established by reference to the individual parts, and what one understands of each individual part is established by reference to the whole [9].

Our approach begins with determining and explaining the meaning of the Japanese version of the smart city. This followed by literature reviews describing the initiatives adopted by the Japanese government and the companies supporting the smart grid concept of going green and the attempts to create a low-carbon society. Then, Japan's energy resources are listed, and its major power generation sources and photovoltaic (PV) power generation development are elaborated on. The efforts are then evaluated by using Japan's smart grid community approach and its smart grid pilot projects layout. Note that the four main stages of this study are supported by three main elements—theory, data and practice—, which serve as a strong reference for the sources obtained. Initiative, investment, and collaboration are the main keywords in our contribution to the literature review, which design to ensure that the sources used are relevant to the topics studied.

## III. CONCEPT

The main objectives of adopting smart grid technology differ by region. Japan's main objective is to achieve a total shift from fossil fuels to renewable energy [10], generating a low-carbon society. Reducing carbon dioxide ($CO_2$) emissions is by no means easy, and is thought to require a large number of combined measures. One such measure is the utilization of renewable energy, and Japan promotes this measure through the development and employment of the natural grid.

In June 2010, the Japanese cabinet adopted a new Basic Energy Plan. This was the third such plan that the government had approved since the passage of the Basic Act on Energy Policy in 2002, and it represented the most significant statement of Japanese energy policy in more than four years, since the publication of the New National Energy Strategy in 2006. Among the targets are a doubling of Japan's energy independence ratio, a doubling of the percentage of electricity generated by renewable sources and nuclear power, and a 30% reduction in energy-related $CO_2$ emissions, all by 2030 [11].

### A. Low-carbon Society

It was believed that a low-carbon society would not be realized without a fundamental shift in energy source use. The Japanese smart grid concept aims to make the best use of local renewable energy with a view to maximizing total efficiency.

The Japanese government is aiming to increase the reliability of the grid system by introducing sensor networks and to reduce opportunity losses by introducing smart meters. The introduction of the smart grid will promote the use of renewable energy by introducing a demand response system. By focusing on electric vehicle (EV) technology, Japan is moving toward introducing charging infrastructure for electric cars [10]. Recently, increasing numbers of PV and wind power plants have been installed across the country as clean energy sources that emit no $CO_2$ [12].

### B. Natural Grid

There is an urgent need to develop and adopt products, processes, and behaviors that will contribute toward more sustainable use of natural resources. In developing new green technologies and approaches, engineers in Japan are finding inspiration from natural products and systems. Since 2010, the Tokyo Electric Power Company (TEPCO) and the Kansai Electric Power Company have been testing the effects of smart meters on load leveling under the Agency for Natural Resources and Energy project.

There are five major elements in a community grid system. The smart office refers to intelligent building design involving cabling, information services, and environmental controls, and envisages a desire for architecture with permanent capacity for EVs for office energy backup. In order to accelerate grid modernization in schools, the Japanese government started to develop a vision of schools that operate by drawing energy supply from PVs; this government program assists in the identification and resolution of energy supply issues and promotes the testing of integrated suites of technologies for schools. The smart house projects relate to smart houses interacting with smart grids to achieve next-generation energy efficiency and sustainability [13], and information and communication technology-enabled collaborative aggregations of smart houses that can achieve maximum energy efficiency. The potential benefits of smart houses include cheaper power, cleaner power, a more efficient and resilient grid, improved system reliability, and increased conservation and energy efficiency. A smart house enables PVs and EVs to stabilize demand and supply fluctuations. Smart factory systems enable full factory integration of the PV cell into the bitumen membrane. The PVs and EVs in a smart factory are supplied to support its production process. Smart stores refer to the charging outlets in parking areas and the deployment of public charging stations for EVs. The arrows indicate that there exist real-time energy flows and information flows in a community grid system. Japan's national grid system will be enhanced by utilizing low-energy sources such as geothermal, hydraulic, battery systems, and nuclear.

The natural grids are expected to play a vital role in making effective use of renewable energy and providing a stable supply of power by controlling the balance between electricity supply and demand by using telecommunications technology.

### C. A Comparison of the Smart Grid Movements in Japan and the US

The energy sources in Japan and the US differ greatly, and the implementation of the smart grid tends to differ between countries, as do the timing and adoption of these technologies [14]. Japan is pushing for advanced integrated control, including demand-side control, so as to be ready to combine unstable power (that is, reliant on weather conditions), such as solar, with its strong foundation of a highly reliable grid, as show in Table I.

With regard to Japan's nuclear contribution to energy supply, as of 2010, Japan's currently operating 54 commercial nuclear reactors have a total generation capacity of 48,847 MW, and about 26% of electricity comes from nuclear power [15]. This compares with 104 licensed-to-operate nuclear power plants operating in 31 states in the US (with 69 pressurized water reactors and 35 boiling water reactors, which generate about 19% of US electrical power [16]. Japan's 10 electric power companies are monopolies, being electric giants vertically integrated in each region. By contrast, there are more than 3,000 traditional electric utilities established in the US, each with an interdependent structure. For this reason, utilities' supply of nuclear energy to customers differs greatly between Japan and the US. In aiming toward a low-carbon society, Japan depends greatly on nuclear power as an energy source, whereas the US, in implementing its smart grid, focuses more on business and infrastructure.

In terms of reliability, Japan already has a highly reliable grid compared with the US, which needs more reliable and distributed networks across the nation to develop its smart grid system. Japan is developing its smart grid at a steady pace and has already been investing in grid projects for almost 20 years; over this period, there have been many developments. With proper security controls, smart grids can prevent or minimize the negative impact of attacks by hackers and thus increase the reliability of the grid, thereby gaining the trust and meeting the satisfaction of users [17].

TABLE I. COMPARISON OF JAPANESE AND US SMART GRID EFFORTS

| Description | Japan | US |
|---|---|---|
| Nuclear as % of all energy sources | 26% | 19% |
| No. of electric power companies | 10 electric power companies (all IOUs) | More than 3,000 traditional electric utilities (IOUs=210, public=2,009, coops=883, federal=9) |
| Design | Vertically integrated in each | Interdependent infrastructure |
| | region | |
| Energy supply | 0.7–25 million customers | A few thousand to more than five million customers |
| Aim | A low-carbon society | Focus on business and infrastructure |
| Reliability | - Japan already has a highly reliable grid<br>- Going for advanced integrated control, including demand-side control, to accommodate unstable renewable power | - Need for highly reliable transmission and distribution networks<br>- Need for demand response for peak shaving and need to avoid additional infrastructure |
| Smart grid focus | - More than $100 billion investment in the 1990s to upgrade generation, transmission, and SCADA network<br>- Last mile and demand-side management (DSM)<br>- Home solar power | - Little investment (approx. $30 billion) in the 1990s into grid<br>- Now working across entire grid for enhancements<br>- Last mile and DSM are also important |
| Cyber security research | Protect smart meters, mutual monitoring, privacy in cloud computing | Grid computer, cryptography security |

Since smart grid cyber security is significantly more complex than the traditional IT security world, Japan focuses on areas of smart grid cyber security concerns beyond smart metering, such as mutual monitoring and privacy in cloud computing. On the other hand, the US is enhancing its smart grid cyber security in the age of information warfare, especially in the area of grid computer and cryptography security.

In conclusion, the US focuses on businesses and infrastructure, whereas Japan is striving to move toward a low-carbon society by developing the smart grid system.

### IV. ENERGY SOURCES

Energy resources in Japan are very limited and for that reason about 97% of oil and natural gas has to be imported; about half of these primary energy sources are converted to electric power; the commercial and residential sectors account for about 27% of total energy consumption, of which space heating and air conditioning account for 24.5% of total household electricity consumption [19].

Japan's energy sources can be categorized into 11 groups: electric power for commercial and industrial use; electric power for residential use; gasoline; kerosene; heavy oil; light oil; city gas; butane gas; propane gas; coal; and coke.

### A. Japan's Major Power Generation Sources

Unlike most other industrial countries, Japan does not have a single national grid, but instead has separate eastern and western grids. The standard voltage at power outlets is 100 V, but the grids operate at different frequencies: 50 Hz in eastern Japan and 60 Hz in western Japan [20]. Japan's major power generation sources are list as below.

#### 1) Hydroelectric Power

This is one of the few self-sufficient energy resources in resource-poor Japan. Hydroelectric power is an excellent source in terms of stable supply and generation cost over the long term [12]. Although steady development of hydroelectric power plants is desired, Japan has used nearly all available sites for the construction of large-scale hydroelectric facilities, and so recent developments have been on a smaller scale. As the gap in demand between daytime and nighttime continues to grow, electric power companies are also developing pumped-storage power generation plants to meet peak demand. The share of pumped-storage generation facilities in total hydroelectric power capacity in Japan is growing year by year. The Okumino Hydroelectric Power Plant functions as a pumped-storage plant; the other one is the Arimine Daiichi Hydroelectric Power Plant [21].

#### 2) Thermal Power

A diverse range of fuels including coal, oil, and liquid natural gas (LNG) are used for the important power-generating role played by thermal power plants. In particular, in response to global environmental concerns, electric power companies are promoting the introduction of LNG-fired plants, as they emit less $CO_2$ and other pollutants. To enhance thermal efficiency further, combined-cycle generating plants with both gas and steam turbines have been installed. As a result, the gross thermal efficiency, or the maximum designed value, now exceeds 50%.

The two major thermal power plants are the Noshiro Thermal Power Plant (coal fired) and the Nanko Thermal Power Plant (LNG fired). Despite its small size, Japan has the third largest geothermal energy potential in the world after the US and Indonesia. However, in terms of harnessing that heat and turning it into power, Japan only ranks eighth, after countries with much smaller populations, such as Iceland and New Zealand. Today, Japan only generates about 0.1% of its electricity in 19 geothermal energy plants, many of which are located in the Tohoku region, where the Fukushima nuclear power plant is located [22].

#### 3) Nuclear Power

Japan's first commercial nuclear power plant started operation in the Ibaraki Prefecture in 1966. As of January 19, 2011, Japan has 54 reactors operating around the country [23], accounting for around one-third of the country's total electric power output. By 2018, the nuclear output share is expected to reach 40%. Currently, there are 3 plants under construction, and another 10 that are in the advanced planning stages.

Establishing a smart grid has been considered problematic in Japan because of the monopoly on electricity supply, and hence, there has been virtually no discussion of the smart grid. Japan's recent tsunami-induced nuclear crisis has, not surprisingly, sent a signal about TEPCO's ability to supply sufficient energy to the nation. The World Nuclear News reported that Japan derives about 30% of its electricity from 54 nuclear reactors, seven of which were down for routine maintenance when the earthquake and tsunami struck on March 11, 2011. Four more reactors were disabled during the disaster and remain unstable, and more than one-third of Japan's nuclear-generated power was unavailable during that period [24]. Three days after the tsunami struck, TEPCO reported that its 12 thermal power units and 22 hydroelectric units had been knocked out by the earthquake. This meant that only 33 GW of capacity were available to meet 37 GW of demand on the day of the earthquake, resulting in power outages to 2.4 million households [25]. This phenomenon has made the government realize that it needs a plan to overcome sudden energy shortages.

### B. PV Power Generation

The Japanese government developed its new policy on PV systems in 2010 based on the principles of energy security, global warming prevention, and an efficient supply of energy to end users across the nation. Its goal, set to be achieved by 2030, is focused on increasing independent energy supplies from 38% to about 70%, and increasing the proportion of electricity generation with zero emissions from 34% to about 70% [26]. This new policy is supported by the government's aim of becoming a leading nation in environmental and energy sustainability through green innovation.

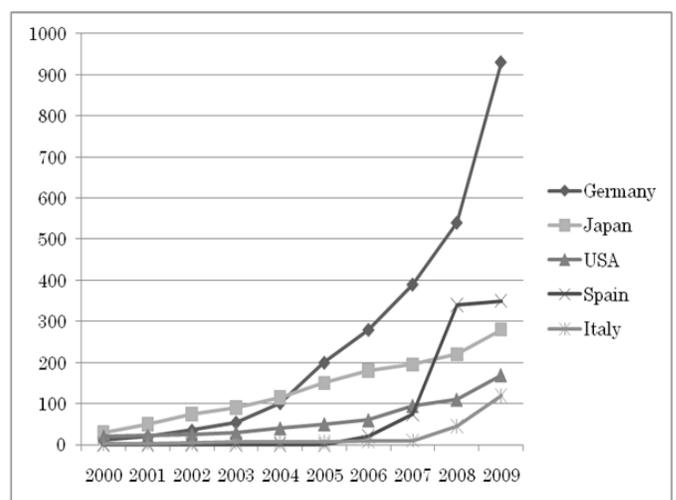

Source: IEA PVPS

Figure 1.   Cumulative Installation of PVs in Japan, Germany, the US, Spain, and Italy

Figure 1 illustrates the cumulative installation of PVs in Japan and four other countries, namely Germany, the US, Spain, and Italy, extracted from data on trends in PV applications. As of 2009, Japan clearly ranks third, lagging far behind Germany. The demand for PV systems in Germany has remained persistently high for a full two years. Spain ranks second and Spanish companies and research centers are taking the lead in the recent revival of concentrated solar power, with expansive banks of solar mirrors being assembled around the country for concentrated solar plants.

In the context of Japan's PV power generation development, the expected change in domestic electricity demand in Japan in relation to the expected installed PV system capacity by 2030, along with the actual cumulative installed PV system capacity as of 2007. Domestic electricity demand is expected to increase sharply between 2010 and 2030. This is correlated with the expected installed PV system capacity, which experienced a sharp upturn in 2010, following the development of the detection of unintentional islanding, and the development of technology that curtails the restriction of PV system output. Expanding installation of PV systems may increase the stability of extra-high voltage transmission systems. At the final stage of PV system development, it is predicted that imbalances between output from PV systems and existing systems may influence frequency on utility systems.

The New Energy and Industrial Technology Development Organization (NEDO) and the European Commission (European Union) will jointly launch a project to develop concentrator PV cells, thus aiming to achieve a cell conversion efficiency of more than 45%, which is the highest efficiency in the world [27].

In addition, Sunetric, Hawaii's largest locally owned and operated solar installer, has donated two solar PV systems to raise funds for two local charities assisting Japan following the tsunami that hit northeast Japan on March 11, 2011. The first is the American Red Cross Hawaii State Chapter and the second is the 'With Aloha' Foundation [28].

In summary, the Japanese government is encouraging further deployment of the conventional installation of residential PV systems for the sake of the PV community. Its current PV communities include Kasukabe and Yoshikawa in Saitama, Matsudo in Chiba, Kasugai in Aichi, Kobe in Hyogo, Tajiri in Osaka, Ota in Gunma, Wakkanai in Hokkaido, Shimonoseki in Yamaguchi, and Kitakyusyu in Fukuoka. Although the national subsidy program for residential PV systems introduced by Japan's Ministry of Economy, Trade, and Industry (METI) was terminated, some local governments have programs for residential PV systems in the regions.

### C.  A Comparison of Japanese and US Energy Sources

Japan lacks significant domestic sources of fossil fuels, except coal, and must import substantial amounts of crude oil, natural gas, and other energy resources, including uranium. In 1990, Japan's dependence on imports for primary energy stood at more than 85%, and the country has a total energy requirement of 428.2 million tons of petroleum equivalent [29]. Figures 2 and 3 compare energy sources in Japan and the US.

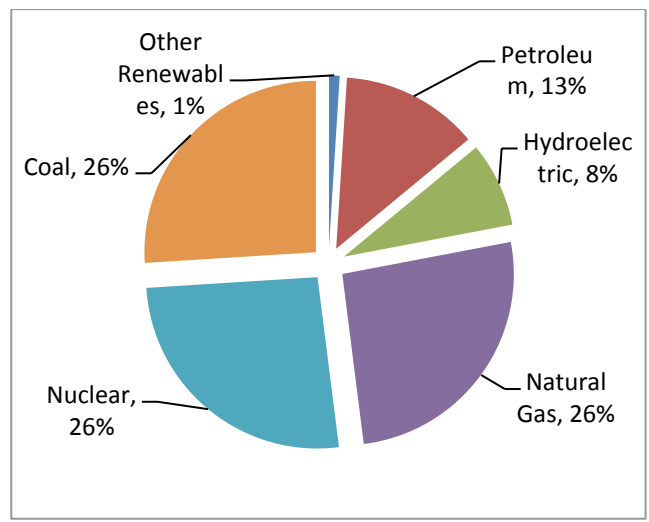

Source: EIA

Figure 2.   Energy Sources in Japan (data of 2007)

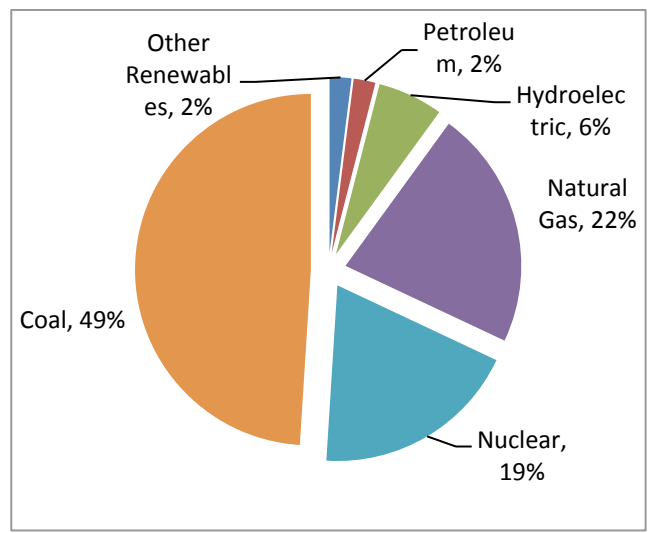

Source: EIA

Figure 3.   Energy Sources in the US (data of 2007).

For Japan, the shares for coal, natural gas, and nuclear energy are similar, and the remaining share is split between hydroelectricity, petroleum, and other renewables. For the US, the large share for coal is followed by natural gas and nuclear, with smaller shares for hydroelectricity, then petroleum. Japan is to be commended for having such a systematic and comprehensive energy planning process. While maintaining its goal of going green, Japan is utilizing more low-carbon energy sources such as geothermal, hydraulic, battery systems, and nuclear as major providers of energy.

## V. JAPAN'S SMART COMMUNITIES

Japan's smart community initiative is based on a systemic approach. There are five identified action items: sharing vision and strategy for smart communities; social experiments for development and demonstration; standardization and interoperability; data-driven innovation and privacy protection; and smart communities for development. To address simultaneously the three Es (environment, energy security, and economy) requires the right mix and match of power sources through renewable and reusable energy (RE) utilizing storage.

TABLE II.    JAPAN'S FUTURE SOCIAL SYSTEM

Source: NEDO

| | *Current period to 2020* | *Period from 2020 to 2030* | *2030 onward* |
|---|---|---|---|
| Relation between regional EMS and entire grid | Solar panel prices will decrease significantly because of the large-scale introduction of panels to houses and commercial buildings. | Because of a decline in PV prices, more PV systems will be installed at houses. | Cost competitiveness of RE will improve as fossil fuel prices increase more than twofold. Use of RE will be prioritized and nuclear power will be used as a base. |
| | Measures will be introduced to maintain the quality of electricity supply while the large-scale introduction of PV systems is conducted mainly for the grid side. Storage cells will be installed at substations. | A regional EMS, which contributes to the effective use of RE generated at houses, will become more important. | An EMS that can provide an optimized balance in terms of economy and security between regional EMS and the grid will be established. |
| | As the regional EMS is further demonstrated, technology and knowledge will be accumulated. | A regional EMS will be achieved as storage cells become cheaper and are further disseminated. | An EMS that creates demand by charging EVs at the time of excessive RE reliance, and supplies energy to the grid at times of high demand, will be used. |
| | The cost of storage cells will decrease because of technology development and demonstration. | Distribution and transmission networks that enable two-way communication between the demand side and the grid side will be actively established. | |
| Houses | Remote reading using smart meters will start. | The home EMS and the regional EMS will be integrated. All power generated at houses will be used optimally. | A fully automated home EMS will be achieved. |
| | The home EMS will be disseminated. Some houses will install home servers. Demand response demonstration will start. | Various services using home servers will be disseminated. | |
| | Demonstration of EVs will start. | EVs will be used for power storage as well. | |

Table II illustrates the future social system at which Japan is aiming, concentrating on the regional energy management system (EMS) and lifestyle changes under such an energy supply structure.

The development of Japan's smart community is divided into three stages: the first is the development plan for the current period up to 2020; the second is the development plan for 2020 to 2030; and the third is the development plan for 2030 onward. With regard to the relation between the regional EMS and the grid, the decrease in solar panel prices following their large-scale introduction will cause more PV systems to be installed at houses, which is expected to create cost competitiveness in RE. The EMS will become more important and will provide an optimized balance in terms of economy and security. When EMS technology and knowledge has accumulated and the cost of storage cells has fallen because of technology development, the distribution and transmission networks that enable two-way communication between the demand and grid sides will be actively established. By that time, the EMS that creates demand by charging EVs at times of excessive reliance on RE, and supplies energy to the grid at times of high demand, will be used.

As for the development of houses, the remote reading of smart meters will start when the home EMS and the regional EMS are integrated and all power generated at houses will be used optimally. At the same time, the home EMS, followed by various services using home servers, will be disseminated. When the demonstration of EVs has started and when EVs are used for power storage, a fully automated home EMS will be achieved. In addition, zero-energy buildings (ZEBs) will be introduced from 2020 to 2030, initially for new public buildings. The introduction of ZEBs is expected to reduce emissions greatly for all new buildings as a group.

The aims of Japan's smart community are: (a) to cut energy costs through competitive advantage and establish a new social system to reduce $CO_2$ emissions; (b) to introduce widespread use of RE; and (c) to facilitate the diversification of power supplier services [30]. This approach is also directed at helping Japanese customers see how cutting their energy costs can give them a competitive advantage.

In addition, a consortium of Japanese companies is preparing a report on the feasibility of smart community development projects in Gujarat, India. According to preliminary estimates by Japanese experts, of the 6.23 million tons of hazardous waste generated in India annually, 22% comes from Gujarat. This report prompted the Gujarat government to sign a memorandum of understanding for developing 'Surat' as an eco-town along the lines of 'Kitakyushu Eco-Town' in Japan and for developing Dahej along the lines of the 'reduce, reuse, and recycle'-oriented environmentally smart community development concepts prevalent in Japan [31].

## VI. Japan's Smart Grid Pilot Projects

On April 8, 2010, four sites were selected from four cities in Japan to run large-scale, cutting-edge pilot projects on the smart grid and smart community (budget request for FY2011: 18.2 billion yen) [25]. The community EMS will be achieved based on a combination of the home EMS, the building EMS, EVs, PVs, and batteries. Not only METI's smart grid-related projects, but also projects in other ministries, such as communications, environment, agriculture, and forestry, will be implemented at these four sites (Figure 4).

Five months later in 2010, the Japan Wind Development Company, the Toyota Motor Corporation, the Panasonic Electric Works Company, and Hitachi Ltd. started a smart grid demonstration project in Rokkasho Village, in the Aomori Prefecture, aiming to verify technologies that allow for the efficient use of energy for the achievement of a low-carbon society [32]. Six months later, the Hawaii–Okinawa Partnership on Clean and Efficient Energy Development and Deployment began, with the aim of helping the two island regions switch from thermal power to renewable energy systems, which are considered crucial for reducing $CO_2$ emissions but whose power supply is unstable [33].

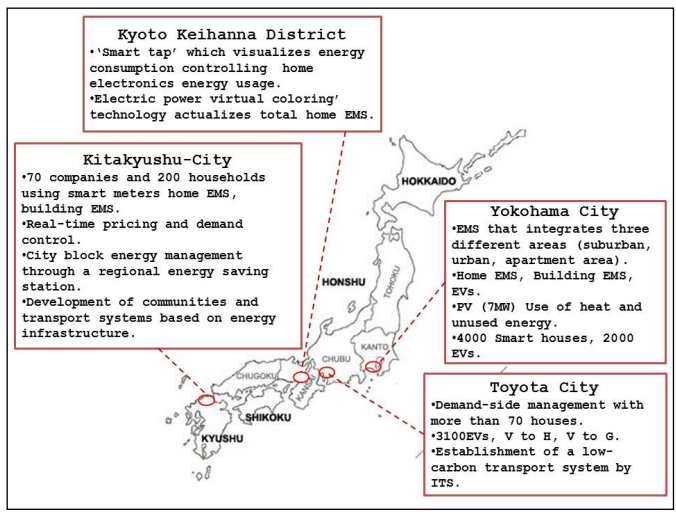

Source: METI

Figure 4.   Energy Sources in the US (data of 2007).

### A. Smart Grid Pilot Project in Yokohama City (Large Urban Area)

There were 900 units of PV systems installed in the so-called progressive city of Yokohama in 2009, and the Japanese government plans to install about 2,000 more 10 years later [34]. The aim of the Yokohama Smart City Project is to build a low-carbon society in a big city, involving 4,000 smart houses. This project is a five-year pilot program being undertaken with a consortium of seven Japanese companies; they are the Nissan Motor Co., Panasonic Corp., Toshiba Corp., TEPCO, the Tokyo Gas Co., Accenture's Japan unit, and Meidensha Corp. The project focuses on the development of the EMS, which integrates the home EMS, the building EMS, and EVs. It is expected to generate PVs with a capacity of 27,000 KW. The

EMS, which integrates suburban, urban, and apartment areas, will have PV use of heat and unused energy.

### B. Smart Grid Pilot Project in the Kyoto Keihanna District (R&D Focus)

The Smart Grid Pilot Project in the Kyoto Keihanna District involves the Kyoto Prefecture, Kansai Electric Power, Osaka Gas, Kansai Science City, Kyoto University, Doshisha, Yamate, the Sustainable Urban City Council, and other local governments and utilities. It makes use of the 'smart tap', which visualizes energy consumption controlling home electronics energy usage. It is also a pilot project to test 'electric power virtual coloring' technology, which actualizes the overall home EMS. This project calls for the installation of PVs in 1,000 houses and an EV car-sharing system. It also studies nanogrid management of PVs and fuel cells in houses and buildings on the visualization of demand. This project grants 'Kyoto eco-points' for the usage of green energy.

### C. Smart Grid Pilot Project in Toyota City (Regional City)

The Toyota Rokkasho Village in the Aomori Prefecture began to experiment with the smart grid in September 2010. The key feature of the project is the pursuit of optimal energy use in living spaces at the community level at the same time as achieving compatibility between environmental preservation and resident satisfaction. This project involved Toyota City, and companies including Toyota Motors, Chubu Electric Power, Toho Gas, Utilities, Denso, Sharp, Fujitsu, Toshiba, KDDI, Circle K Sunkus, Mitsubishi Heavy Industries, and Dream Incubator. The project focuses on the use of heat and unused energy as well as electricity. It has a demand response with more than 70 homes and 3,100 EVs. Through this project, houses that contain an IT network of electrical appliances and other household equipment, solar panels, household storage batteries, onboard automobile storage batteries, and other devices, can develop household power leveling and optimized energy usage. As of June 2011, model smart houses in the Higashiyama and Takahashi districts of Toyota City for testing EMSs had been completed successfully and had begun trial operations under the Verification Project for the Establishment of a Household and Community-Based Low-Carbon City in Toyota City, Aichi Prefecture [35].

### D. Smart Grid Pilot Project in Kitakyushu City (Industrial City)

This project involves 46 companies and organizations, including the Kitakyushu City Government, GE, Nippon Steel, IBM Japan, and Fuji Electric Systems. It focuses on real-time management in 70 companies and 200 houses. Energy management will be controlled by the home EMS and the building EMS. The pilot project study of the energy system integrates demand-side management and the high-energy system. The Kitakyushu Hibikinada area is promoting low carbon emissions, recycling, and nature coexistence in a balanced manner. This project implements various demonstrations, including communications, urban planning, a transportation system, and lifestyle, with an emphasis on the demonstration of energy projects such as electric power.

Implementation within five years, from FY2010 to FY2014, involving the operation of 38 projects, is worth 16.3 billion yen [36].

The Kitakyushu Smart Community Project focuses on the development of technologies and systems related to a smart grid with an eye on international standardization and the expansion of international business, and the presentation of new urban planning for a smart city, by developing various human-friendly social systems compatible with next-generation traffic systems and an aging society. In addition, a recycling community will be constructed in which all sorts of waste will be used as raw materials for other industrial fields to eliminate waste and move toward a zero emissions community.

### E. Smart Grid Demonstration Project in Rokkasho Village

In September 2010, the Toyota Motor Corp. began a two-year project with Japanese Wind Development, Panasonic Electric Works, and Hitachi Ltd. in the village of Rokkasho, Aomori Prefecture, where wind power stations with large-capacity batteries were built several years ago [8]. The project involves a so-called smart grid village composed of six 'smart houses' equipped with automatic electricity control systems, eight Toyota Prius plug-in hybrid vehicles, and a battery system, all powered exclusively by renewable energy sources and detached from the national electricity grid. Families of the employees of the corporations participating in the project reside in these smart houses, where they go about their normal lives [37].

In addition, an experimental situation has been created in isolation from the external power grid, where approximately eight kilometers of private distribution line has been laid between the Rokkasho-mura Futama Wind Power Station and where the smart houses stand. The station is outfitted with 34 units of 1,500 KW windmills with a total capacity of 51,000 KW, and is equipped with large-capacity network-attached storage batteries of 34,000 KW. The objective of the experiment is to examine such factors as changes in electricity usage in different seasons and at different times of day, and investigate trends in electricity usage based on different family configurations. The experiment will help create a system that efficiently balances electricity supply and demand.

### F. Smart Grid Trail Project in Okinawa

The Okinawa Electric Power Company has begun operating a smart grid to control the supply of renewable-energy-derived electricity for the 55,000-strong population of the remote Okinawa Prefecture island Miyako-jima. The project is part of the Hawaii–Okinawa Partnership on Clean and Efficient Energy Development and Deployment, an agreement between the US Department of Energy, METI, the State of Hawaii, and the Prefecture of Okinawa, which was signed in June 2010.

The Hawaii–Okinawa partnership is intended to foster the development of clean and energy-efficient technologies needed to achieve global energy security and meet climate change challenges. Japan and the US designated Hawaii and Okinawa as the representatives for this groundbreaking partnership because of their demonstrated leadership and experience in clean energy and energy efficiency. The trials started in October 2010. The infrastructure links the existing power grid to a four MW solar power plant and a sodium sulfide battery complex capable of storing four MW of power. Some lithium ion batteries have also been installed. In addition, the system also controls power from existing 4.2 MW wind farms situated on Miyako-jima. Okinawa Electric spent 6.15 billion yen (US$75.8 million) on the infrastructure, two-thirds of which was subsidized by the national government [33]. The programs will help the two island regions switch from thermal power to renewable energy systems, which are considered crucial for reducing $CO_2$ emissions, but whose power supply is unstable.

Each of the smart grid pilot and demonstration projects has its own challenges. Building smart grids requires meeting the requirements for the electricity supply, including the power sources and transmission lines, and the communications infrastructure of each specific country and region, as well as introducing such elements as renewable energy generation facilities, EVs and plug-in hybrid vehicles, storage batteries, Eco Cute electric water heating and supply systems, and heat storage units.

Currently, possibilities in new electricity distribution methods and vast advances in information and communications technology are raising the prospect of a shift from today's conventional supply–demand adjustment approach to one that optimizes both supply and demand. The Japanese government is strongly promoting the efficient use of energy by developing smart grid pilot projects and the promotion of the natural grid community.

### VII. JAPANESE COLLABORATIONS ON SMART GRID PROJECTS

### A. Smart Grid Project in Hawaii

A project supported by Japan's NEDO, in cooperation with the State of Hawaii, the Hawaiian Electric Company, the University of Hawaii, and Pacific Northwest National Laboratory, whose involvement is based on the Japan–US Clean Energy Technologies Action Plan, was started in November 2009 [38]. Hitachi Ltd., Cyber Defense Institute Inc., the JFE Engineering Corporation, Sharp Corporation, Hewlett–Packard Japan Ltd., and Mizuho Corporate Bank Ltd. were among those selected as contractors for a joint Japan–US collaboration supporting a smart grid project on the Hawaiian island of Maui, which will serve as the project site. A feasibility study is expected to be completed by the middle of September 2011 [39]. Subject to the results of the feasibility study, the project is expected to be implemented by the end of March 2015.

While appearing to be 'state-of-the-art' technological marvels, smart grids expose energy production and distribution to higher levels of security vulnerability than ever before. In considering this matter, the project will incorporate the installation of smart controls in the Kihei area on Maui at the regional and neighborhood levels to improve the integration of variable renewable energy resources, such as PV systems [40]. Installation of the smart grid technology is expected to begin in late 2012, with the project becoming operational in 2013. The demonstration project is scheduled to run from 2013 to 2015. This project may be useful for the design of future micro-grids that will provide secure backup and UPS services to distributed energy residences and light industry. Independent, distributed energy appliances in homes and businesses guarantee the highest possible level of security and reliability in a national power system.

### B. Smart Grid Pilot Project on the Island of Jeju

Large electronics conglomerates in Japan and Korea, such as Sharp, Panasonic, Samsung, LG, SK Telecom, and KT, are building a domestic smart grid pilot on the island of Jeju, which is south of Seoul in South Korea [41].

In March 2011, IBM announced that two new utilities had joined its Global IUN Coalition, a group of utility companies designed to further the adoption of smarter energy grids around the world: TEPCO from Japan and KEPCO from Korea. These two companies are in charge of the world's largest comprehensive smart grid test bed in Jeju Island, which brings together smart technologies in the areas of generation, power grids, electrical services, buildings, and transportation [42].

### C. Smart Grid Project in New Mexico

Toshiba, Kyocera, Shimizu, the Tokyo Gas Company, and Mitsubishi Heavy Industries will spend $33.4 million on a smart grid project at Los Alamos and Albuquerque, New Mexico [30]. NEDO will participate in research at Los Alamos and Albuquerque and in collective research on the overall project. Toshiba will install a one MW storage battery at the Los Alamos site, while Kyocera and Sharp will test smart homes, energy management, and load control technologies.

The Microgrid Demonstration Project in Los Alamos involves concentrated PV energy generation and the installation of power storage cells on distribution lines of 2 to 5 MW. In addition, absorption experiments on PV output fluctuations will be conducted using PV-induced efficiencies obtained by changing grid formation, and a distribution network with high operability will be installed and demonstrated by introducing smart distribution equipment. The smart house is intended to maximize demand response by using a home EMS and a demonstration will be carried out to verify its effectiveness relative to an ordinary house. The micro-grid demonstration in commercial areas of Albuquerque focuses on demonstrating the demand response by using facilities in industrial and commercial buildings. The move is prompted by the aim of catching up with the US, which has taken the lead in developing technological global standards [41]. It is also intended to evaluate smart grid technology from Japan and the US based on research results obtained at the five demonstration sites of the New Mexico project [30].

### D. ZEBs in Lyon in France

NEDO is holding discussions with Grand Lyon, the second largest city in France, to introduce Japanese leading-edge technologies for ZEBs in France and to establish an EV-charging infrastructure coinciding with the Lyon Confluence urban development project in Lyon [30].

### VIII. MAJOR SMART GRID AWARENESS AND ACTIVITIES PROMOTION BODIES IN JAPAN

One important initiative is having members engage in smart grid promotional activities that complement activities in existing organizations and groups that currently influence and shape policy, architecture, standards, and traditional utility operations [43]. A few major smart grid awareness and activities promotion bodies and associations affirm the successful implementation of smart grids in Japan and raise people's awareness through education.

### A. The Japan Smart Community Alliance (JSCA)

The JSCA is a member of the Global Smart Grid Federation [44], which aims to promote public–private cooperative activities relating to the development of smart communities by tackling common issues such as dissemination, deployment, and research on smart grid standardization. The JSCA has members from the electric power, gas, automobile, information and communications, electrical machinery, construction, and trading industries as well as from the public sector and academia. Four working groups have been established at a practical level for discussion and deliberation in order to facilitate the JSCA's activities. The four working groups are the International Strategy Working Group, the International Standardization Working Group, the Roadmap Working Group, and the Smart House Working Group. Each group supports smart grid development in Japan.

### B. METI

METI is reviewing closely Japan's PV/CSP technologies and programs to support smart grid development in Japan.

### C. Democratic Party of Japan (DPJ)

The DPJ is promoting the development and diffusion of smart electricity grid technologies.

### D. NEDO

While NEDO is committed to contributing to the resolution of energy and global environmental problems and further enhancing Japan's industrial competitiveness, it strongly supports numerous smart grid research and development projects. NEDO aims to develop the world's most efficient concentrator PV cells.

In addition, Japan is promoting smart grids by conducting discussions and undertaking projects in an integrated manner with the participation of various stakeholders [45]. Two months ago, nine Japanese companies - Shimizu Corporation, Toshiba Corporation, Sharp Corporation, Meidensha Corporation, Tokyo Gas Co., Ltd., Mitsubishi Heavy Industries, Ltd., Fuji Electric Co., Ltd., Furukawa Electric Co., Ltd. and The Furukawa Battery Co., Ltd. launched a demonstration study for the Albuquerque Business District Smart Grid Demonstration Project consigned to them by the New Energy and Industrial Technology Development Organization (NEDO), to be carried out as part of its Japan-U.S. Collaborative Smart Grid Demonstration Project, took will took place form March 2012 to March 2014 [46].

## IX. CONCLUSION

Smart technologies improve human ability to monitor and control the power system. Smart grid technology helps to convert the power grid from static infrastructure that is operated as designed to flexible and environmentally friendly infrastructure that is operated proactively. This is consistent with the Japanese government's goal of creating a low-carbon society, maximizing the use of renewable energy sources, such as photovoltaic and wind power. Nevertheless, public–private sector cooperation across various industries is necessary to establish smart communities.

This paper provided sufficient information for the reader to understand the Japanese concept of the smart grid and the government's associated strategy and the significance of the government's contribution to Japan's energy supply capacity, without documenting full case studies in detail.

Japan chosen to be at the boundary at some time in the duration of the smart grid revolution because they already have had a reliable grid system. Recent the Japanese association is bringing up large-scale grids that deal with the power like how the internet does with the information data.

This paper painted a conceptual picture of Japan's smart community initiatives and its investment in and collaboration on smart grid pilot projects. With emphasis on Japanese initiatives, investment, and collaboration, comparison tables, figures, and graphs relating to smart grid developments were used to enable understanding of the issues. Although the Asia Pacific region is quickly catching up in smart grid developments and adoption, because the energy sources of different countries vary significantly, the methods and timing with which countries adopt this technology differ. Japan is currently focusing on last mile and demand-side management and home solar power. Researchers have started to address challenges caused by large-scale solar power generation connected to the power grid as well as information security issues. Because the smart grid remains a novel field of study in Japan, it has great potential for further research.

## ACKNOWLEDGMENT

This study was supported by the Meiji University Global COE Program "Formation and Development of Mathematical Sciences Based on Modeling and Analysis", Meiji Institute for Advanced Study of Mathematical Sciences (MIMS), and the Japan Society for the Promotion of Science (JSPS).

## REFERENCES

[1] Amy Poh Ai Ling, Mukaidono Masao, "Grid Information Security Functional Requirement Fulfilling Information Security of a Smart Grid System", International Journal of Grid Computing & Applications (IJGCA), Vol. 2, No. 2, June 2011, pp. 1–19.

[2] Tomoko A. Hosaka, "Japan Creating 'Smart City' of the Future", October 2010, pp. 124–135.

[3] Tatsuya Shinkawa, "Smart Grid and Beyond", New Energy and Industrial Technology Development Organization (NEDO), 2010.

[4] Hiroko Nakata, "Smart Grid Pursuit Slow Off Mark", Smart Grid, The Japan Times, January 2011.

[5] Ahmad Faruqui, Ryan Hledik, Sanem Sergici, "Piloting the Smart Grid", The Electricity Journal, Vol. 22, Issue 7, August–September 2009, pp. 55–69.

[6] Feilong Tang, Minglu Li, Joshua Zhexue Huang, "Real-Time Transaction Processing for Autonomic Grid Applications", Journal of Engineering Applications of Artificial Intelligence, Vol. 17, Issue 7, October 2004, pp. 799–807.

[7] Amy Poh Ai Ling, Mukaidono Masao, Sugihara Kokichi, "The Natural Grid Concept and the Strategy of Asia's Energy-Balance Pioneer", The Ninth International Conference on Advances in Mobile Computing and Multimedia, submitted.

[8] Heinz K. Klein, Michael D. Myers, "A Set of Principles for Conducting and Evaluating Interpretive Field Studies in Information Systems", MIS Quarterly, Vol. 23, Issue 1, May 1999, pp. 67–94.

[9] Ramberg Bjorn, Kristin Gjesdal, "Hermeneutics", Stanford Encyclopedia of Philosophy, Library of Congress Data, November 2005.

[10] Shinsuke Ito, "Japan's Initiative on Smart Grid–A Proposal of 'Nature Grid'", Information Economy Division, Ministry of Economy Trade and Industry, December 2009, available at http://documents.eu-japan.eu/seminars/europe/other/smart_grid/presentation_ogawa.pdf

[11] John S. Duffield, Brian Woodall, "Japan's New Basic Energy Plan", Energy Policy, Vol. 39, Issue 6, June 2011, pp. 3741–3749.

[12] "Electricity Review Japan", The Federation of Electric Power Companies of Japan, 2011.

[13] Anke Weidlich, "Smart House / Smart Grid", ICT Challenge 6: Mobility, Environmental Sustainability and Energy Efficiency, Smart Houses Interacting with Smart Grids to Achieve Next-Generation Energy Efficiency and Sustainability Project ICCS-NTU,. December 2009.

[14] Kelly McGuire, "The Smart Grid Movement: Japan vs. U.S.", TMCnet News, Technology Marketing Corporation, January 2010, available at http://smart-grid.tmcnet.com/topics/smart-grid/articles/73301-smart-grid-movement-japan-vs-us.htm

[15] "Nuclear Power Plants in Japan", Nuclear Power Generation, Information Plaza of Electricity, The Federation of Electric Power Companies of Japan, June 2011.

[16] "Power Reactors", Nuclear Reactors, United States Nuclear Regulatory Commission (U.S. NRC), May 2011, available at http://www.nrc.gov/reactors/power.html

[17] Amy Poh Ai Ling, Masao Mukaidono, "Selection of Model in Developing Information Security Criteria on Smart Grid Security System", Smart Grid Security and Communications, The Ninth International Symposium on Parallel and Distributed Processing with Applications (ISPA), No. 108, May 2011, Korea.

[18] Ryusuke Masuoka, "Smart Grid: Japan and US", Fujitsu Laboratories of America Inc., January 21, 2010.

[19] Japan Energy Conservation Handbook, 2004–2005, available at http://www.eccj.or.jp/databook/2004-2005e/index.html

[20] Electricity in Japan, Japan Guide, available at http://www.japan-guide.com/e/e2225.html

[21] "Profile of Japan's Major Power Generation Sources, Energy and Electricity, Information Plaza of Electricity", The Federation of Electric Power Companies of Japan, available at http://www.fepc.or.jp/english/energy_electricity/electric_power_sources/index.html

[22] Tomoko A. Hosaka, "Japan Creating 'Smart City' of the Future", Associated Press Japan, October 2010.

[23] "Nuclear Power Plants, World-Wide", European Nuclear Society, available at http://www.euronuclear.org/info/encyclopedia/n/nuclear-power-plant-world-wide.htm

[24] "Rolling Blackouts as Japanese Efforts Continue", Regulation and Safety, World Nuclear News, March 2011.

[25] Phil Carson, "Smart Grid More Attractive, Post-Japan?", Intelligent Utility, Mar. 2011, available at

http://www.intelligentutility.com/article/11/03/smart-grid-more-attractive-post-japan

[26] Masaya Yasui, "Japanese Policy on Photovoltaic Systems", ANRE, METI, Japan, October 2010, pp. 4–7.

[27] "Aiming at Developing the World's Highest Efficiency Concentrator Photovoltaic Cells", New Energy and Industrial Technology Development Organization, European Commission, European Union, May 2011, available at http://www.nedo.go.jp/content/100147758.pdf

[28] "Sunetric Offers Free Solar PV Systems to Raise Money for Japan", Cleantech News, April 2011.

[29] Martyn Williams, "A Legacy from the 1800s Leaves Tokyo Facing Blackouts", Computerworld, Retrieved March 2011.

[30] Hiroshi Watanabe, "Smart Community Activities in Japan", Korea Smart Grid Week, November 2010, available at http://www.ksgw.or.kr/down/pr/KSGW_11_HiroshiWatanabe(101110).pdf

[31] "Kitakyushu City's Challenge Toward a Low-Carbon Society", Kitakyushu Smart Community Creation Project, Green Frontier, 2011, available at http://www.challenge25.go.jp/roadmap/media/s5_en.pdf

[32] "Four Companies Start Smart Grid Demonstration Project in Rokkasho", Technology, Japan Today, September 2010.

[33] "Hawaii–Okinawa Partnership on Clean and Efficient Energy Development and Deployment", METI, June 2011.

[34] "Smart Grid of the Future Urban Development", Japanese Nikkan Kogyo Shimbun, June 2010.

[35] "Toyota City Low-Carbon Project Model Homes Completed", Asahi, June 2011.

[36] "Japan to Help Gujarat in Smart Community Project", Rediff, January 2011,.

[37] "Living Off the Grid", Cover Story, Highlighting Japan, February 2011, available at http://www.gov-online.go.jp/pdf/hlj/20110201/12–13.pdf

[38] "U.S. and Japan Companies Collaborate on Smart Grid Project in Hawaii", Sharp Corporation, May 2011, Press Releases, available at http://sharp-world.com/corporate/news/110517.html

[39] "US and Japan Collaborating on Smart Grid Project in Hawaii; EV Operation and Charging, Including Grid-Balancing Services", Green Car Congress, Energy, Technologies, Issues and Policies for Sustainable Mobility, May 2011, available at http://www.greencarcongress.com/2011/05/hitachi-20110518.html

[40] "US and Japan Work on Maui Smart Grid for Electric Vehicles", Sustainable Transport News, Brighter Energy News, May 2011.

[41] Katie Fehrenbacher, "The New Smart Grid Players: Korea, Japan, China, Oh My!", GigaOM, WordPress, February 2010.

[42] "Japanese, Korean Utilities Join Smart-Grid Coalition", Greenbang, May 2011, available at http://www.greenbang.com/japanese-korean-utilities-join-smart-grid-coalition_16971.html

[43] Guido Bartels, "Smart Grid's Progress Extends Beyond the Boundaries of Countries", Special Interview–Part 5 of 5, Guido Bartels, Chairman of Grid Wise Alliance, Talks About "Smart Grid", Renesas, October 2010.

[44] "Japan Smart Community Alliance", June 2011, available at http://www.globalsmartgridfederation.org/japan.html

[45] Shinsuke Ito, "Japan's Initiative on Smart Grid–A Proposal of 'Nature Grid'", Information Economy Division, Ministry of Economy Trade and Industry, December 2009.

[46] "Nine Japanese Companies Launch Japan-U.S. Collaborative Smart Grid Demonstration Project in Business District of Albuquerque, New Mexico", The Wall Street Journal, pp. 1-2, May 2012.

AUTHORS PROFILE

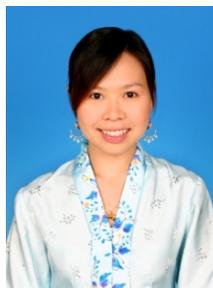

**Amy Poh Ai Ling** received her BBA and MSc from National University of Malaysia (UKM). She received her PhD in Mathematical Sciences from Meiji University. She was awarded Role Model Student Award (2003) and Excellent Service Award (2010) from UKM, and Excellent Student Award (2012) from Meiji University. She worked at Sony EMCS and Erapoly Sdn. Bhd. She is currently a postdoctoral affiliate with Meiji Institute for Advanced Study of Mathematical Sciences as JSPS Research Fellow. She has an enthusiasm for statistical calculation, smart grid and safety studies.

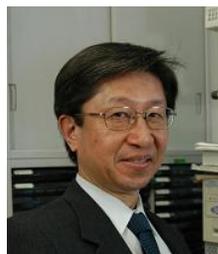

**Professor Sugihara Kokichi** received his Master's and Dr. Eng. degrees from University of Tokyo. He worked at Electrotechnical Laboratory of the Japanese Ministry of International Trade and Industry, Nagoya University and University of Tokyo before joining Meiji University. His research area is mathematical engineering, including computational geometry, computer vision, computer graphics and robust computation. He is currently the leader of CREST research project of Japan Science and Technology Agency on "Computational Illusion".

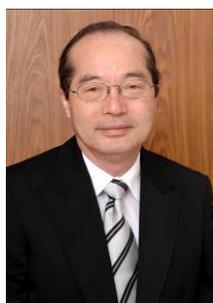

**Professor Mukaidono Masao** served as a full-time lecturer at Faculty of Engineering, Department of Electrical Engineering in Meiji University from 1970. Even since then, he was promoted to Assistant Professor on 1973 and as a Professor on 1978. He contributed as a researcher in an Electronic Technical Laboratory of the Ministry of International Trade and Industry (1974), Institute of Mathematical Analysis of Kyoto University (1975) and as a visiting researcher at University of California in Berkeley (1979). He then became the Director of Computer Center (1986) and Director of Information Center (1988) in Meiji University. At present, he is a Professor and Dean of the School of Science & Technology, Meiji University. He is also the honourable Councillor of Meiji University.